\pdfoutput=1
\documentclass[graybox,hidelinks]{svmult}

\usepackage{mathptmx}       
\usepackage{helvet}         
\usepackage{courier}        
\usepackage{type1cm}        
\usepackage{makeidx}         
\usepackage{graphicx}        
\usepackage{multicol}        
\usepackage[bottom]{footmisc}

\usepackage{xcolor}
\usepackage{framed}

\usepackage{hyperref}
\usepackage{xcolor}

\newcommand{\orcid}[1]{\href{https://orcid.org/#1}{\includegraphics{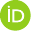}}}

\begin{document}

\hypersetup{pdfauthor={Daniel Weiskopf},pdftitle={Vis4Vis: Visualization for (Empirical) Visualization Research}}

\author{Daniel Weiskopf}

\title*{Vis4Vis: Visualization for (Empirical) Visualization Research}
\institute{Daniel Weiskopf \orcid{0000-0003-1174-1026} \at VISUS, University of Stuttgart, Germany\\ \email{weiskopf@visus.uni-stuttgart.de}\\
ORCID iD: \url{https://orcid.org/0000-0003-1174-1026}\\[2ex]
This is a preprint of a chapter for a planned book that was initiated by participants of the Dagstuhl Seminar 18041 (``Foundations of Data Visualization'') and that is expected to be published by Springer.  The final book chapter will differ from this preprint.}

\renewenvironment{leftbar}[1][\hsize]
{%
    \def\FrameCommand
    {%
        {\color{orange}\vrule width 3pt}%
        \hspace{0pt}
    }%
    \MakeFramed{\hsize#1\advance\hsize-\width\FrameRestore}%
}
{\endMakeFramed}

\maketitle

\abstract{Appropriate evaluation is a key component in visualization research. It is typically based on empirical studies that assess visualization components or complete systems. While such studies often include the user of the visualization, empirical research is not necessarily restricted to user studies but may also address the technical performance of a visualization system such as its computational speed or memory consumption. Any such empirical experiment faces the issue that the underlying visualization is becoming increasingly sophisticated, leading to an increasingly difficult evaluation in complex environments. Therefore, many of the established methods of empirical studies can no longer capture the full complexity of the evaluation. One promising solution is the use of data-rich observations that we can acquire during studies to obtain more reliable interpretations of empirical research. For example, we have been witnessing an increasing availability and use of physiological sensor information from eye tracking, electrodermal activity sensors, electroencephalography, etc.  Other examples are various kinds of logs of user activities such as mouse, keyboard, or touch interaction. Such data-rich empirical studies promise to be especially useful for studies in the wild and similar scenarios outside of the controlled laboratory environment. However, with the growing availability of large, complex, time-dependent, heterogeneous, and unstructured observational data, we are facing the new challenge of how we can analyze such data. This challenge can be addressed by establishing the subfield of {\em visualization for visualization (Vis4Vis)}: visualization as a means of analyzing and communicating data from empirical studies to advance visualization research.}

\author{}
\section{Introduction}
\label{sec:introduction}

This position statement primarily focuses on empirical studies with user involvement but also touches other empirical studies that may collect data from technical performance benchmarks to assess the computational characteristics of a visualization system. 

I argue that we need to establish a new subfield to address the challenges of empirical evaluation in visualization research: 


\leftbar
\mbox{}\\[1ex]
\bf \mbox{}~~~We need {\em visualization for visualization (Vis4Vis)}.\\[0ex]
\mbox{}
\endleftbar

\noindent
The underlying problem is the difficulty in performing an appropriate evaluation for complex visualization systems. For these, many of the traditional approaches to empirical research adopted from other fields cannot be used directly. 
Section~\ref{sec:background} provides background references that discuss various aspects of the underlying problems, methodological challenges, and possible solutions.

I argue that one promising route is to use as much information as possible from empirical studies. Unfortunately, many of the traditional methods for user studies and other empirical research in visualization come from other fields and earlier times in which there was much less data accessible from studies. One example of such data that is still underutilized in visualization research is gaze data from eye tracking experiments. Section~\ref{sec:eyetracking} discusses examples of eye tracking in visualization research in more detail. However, there are many other potential sources of sensor data that could be collected. Several of these examples rely on physiological sensors, often in the context of work on human-computer interaction (HCI): electroencephalography (EEG)~\cite{Anderson:2011:USV} and, in general, the use of brain-computer interfaces (BCIs) and EEG for interaction~\cite{Gurkok:2012:BCI}, pervasive BCI~\cite{Peck:2010:YBY}, near-infrared spectroscopy (NIRS)~\cite{Hirshfield:2011:TYB,Strait:2014:RNB},  functional magnetic resonance imaging (fMRI)~\cite{Cui:2011:QCN}, or the combination of 
several physiological sensors to characterize emotions~\cite{Wagner:2005:PSE}  
or investigate interfaces~\cite{Prendinger:2005:UHP}.

However, data is not restricted to coming from physiological sensors. For example, logging user activities with the visualization interface, based on recording mouse, keyboard, touch, or other ways of interaction, can provide a detailed and rich source of highly relevant information~\cite{Vuillemot:2016:LIV}.
 Other examples are video and audio recordings during user studies that can serve as a basis for think-aloud protocol analysis~\cite{Ericsson:1993:PAV}. 

Overall, technological advances for various kinds of sensors and other data sources have made it easy and cost-effective to capture largely increasing amounts of data for empirical visualization research. And with further progress in technology, in particular, for non-stationary or wearable devices for visualization and user studies, we will see even more diverse types of user studies in visualization research. A recent trend in the visualization community addresses immersive analytics~\cite{Marriott:2018:IA}, which will lead to the problem of evaluating visualizations in the context of virtual reality or augmented reality.

With the challenges of empirical research for complex visualizations on the one hand, and opportunities that come with advanced data acquisition on the other hand, we will have to rethink how we can conduct, evaluate, and report empirical studies. With this text, I focus on the issues related to data analysis for the evaluation and reporting of the results of studies based on large, complex, time-dependent, heterogeneous, and unstructured observational data. I argue that visual data analysis and communication is a promising approach to address these issues. Accordingly, I will discuss opportunities and open questions for visualization research. 
My proposal for the need for {\em Vis4Vis}, especially in the context of empirical visualization research, extends my position statement that I gave as part of the panel discussion at the 2016 Workshop on Beyond Time And Errors: Novel Evaluation Methods For Visualization (BELIV).\footnote{Panel ``On the Future of Evaluation and BELIV'' with panelists Daniel Weiskopf, Laura McNamara, Mark Whiting, Niklas Elmqvist, and Tamara Munzner, BELIV 2016 (Workshop on Beyond Time And Errors: Novel Evaluation Methods For Visualization) at IEEE VIS 2016. \url{https://beliv-workshop.github.io/2016/schedule.html}}

\section{Background of Empirical Studies}
\label{sec:background}

The relevance of empirical studies for evaluation, especially user-oriented evaluation, is well accepted by the visualization research community. 
In general, there are many well-established approaches to empirical studies for visualization and visual analytics~\cite{Carpendale:08:EIV,Plaisant:2004:CIV,Wijk:2013:ECV}. Tory~\cite{Tory:14:USV} provides a recent overview and categorization of user study approaches, covering various quantitative and qualitative methods. Freitas et al.~\cite{Freitas:14:HIC} discuss a user-centered perspective on evaluation. 
There are also examples in which different types of study methods are combined, including the combination of usability metrics and eye tracking \cite{Coltekin:2009:EEI}.

Evaluation methodology is the special focus of the series of BELIV Workshops, which investigate approaches beyond the traditional user performance measures of completion time and accuracy. Therefore, many BELIV Workshop papers address topics relevant to this text. For example, Elmqvist and Yi~\cite{Elmqvist:12:PVE} describe a collection of patterns for evaluation, Ellis and Dix~\cite{Ellis:06:EAU} provide an explorative analysis of user studies, Lam and Munzner~\cite{Lam:08:IUQ} discuss quantitative empirical studies in the context of meta analysis, and Anderson~\cite{Anderson:12:EVU} employs cognitive measures for evaluation.

However, the above papers do not focus on empirical studies that use rich sets of observations. However, Kurzhals et al.~\cite{Kurzhals:2014:EVA,Kurzhals:2016:ETE} consider this approach as critical for future and improved evaluation methods for visual analytics. They especially focus on the combination of eye tracking information with traditional task performance indicators, but they also discuss the issue of data fusion integrating further time-oriented data acquired during an empirical study. One example is the combination of eye tracking and interaction logs \cite{Blascheck:2014:TAE}. Kurzhals et al.~\cite{Kurzhals:2014:EVA} call for exploratory data analysis and hypothesis building to address the difficult analysis questions that come with  complex data. In follow-up work, Kurzhals et al.~\cite{Kurzhals:2016:TBV} adopt the perspective of analysis tasks on eye tracking data, with a respective overview of such tasks. 

A further step in the direction of integrating different data sources from empirical research into an interactive visual analysis approach was taken by Blascheck et al.~\cite{Blascheck2016:VA2,Blascheck:2016:TUB}: they describe how visual analytics methods can be used to evaluate visual analytics systems, for example, by including think-aloud protocol analysis, eye tracking information, or interaction data from the same experiment. Blascheck et al.~\cite{Blascheck:2016:VAC} enrich this approach by integrating visual data analysis and coding of user behavior.

I argue to follow-up and extend this direction of advanced visualization methods for analyzing complex and rich data sources. This will become particularly relevant for studies that address more complex research questions than in traditional, quite focused, and restricted laboratory studies. A trend in HCI and other communities tries to address realistic scenarios by adopting research in the wild \cite{Crabtree:2013:ISI}, following early work on cognition in the wild from the perspective of anthropology \cite{Hutchins:1995:CW,Lave:1988:CP,Suchman:1987:PSA}. A related evaluation need has been identified in the visualization community by Lam et al.~\cite{Lam:2012:ESI} and Isenberg et al.~\cite{Isenberg:2013:SRP}. They discuss scenarios that go beyond traditional user experience, user performance, or (technical) algorithm performance, for example, how we can evaluate communication through visualization,  visual data analysis and reasoning, or collaborative visual data analysis. I am convinced that the visualization of data-rich recordings will be especially useful for empirical research in such areas.

\section{Example: Eye Tracking Studies and Evaluation}
\label{sec:eyetracking}

Let us use eye tracking studies as one example of experimental research with data-rich observations. Gaze is a highly relevant source of data for empirical visualization research because it provides quite accurate and fast information that can be useful to understand attention, reading patterns, and the like. Even though there is not always a direct interpretation of eye tracking data \cite{Kim:2012:DET}, most studies can be set up in a way that eye tracking provides informative feedback if it is used with the right study design and interpretation of results \cite{Goldberg:10:CIG}. Eye tracking might even be an alternative way to measure indicators of insight \cite{North:2006:TMV}.
Background on eye tracking is described in the books by Duchowski~\cite{Duchowski:2007:ETM} and Holmqvist et al.~\cite{Holmqvist:2011:ETC}.

This section focuses on eye tracking for user studies and how we can visually analyze gaze information acquired in such studies. There are other, yet related applications of eye tracking: For example, gaze can serve as a basis for interaction techniques \cite{Jacob:2003:ETH}, eye movements can be employed for activity recognition \cite{Bulling:2011:EMA,Fathi:2012:LRD},  
eye tracking can help identify tasks and abilities of users of information visualizations \cite{Steichen:2013:UAI}, and it can be used to  improve interactive visualization by recommendations built on inferred user interest \cite{Shao:2017:VEL,Silva:2018:LEG} and by adaptive interfaces based on the recognition of user tasks and intent \cite{Silva:2019:ETS}.
 
\begin{figure}[t]
    \centering
        \includegraphics[width=0.95\textwidth]{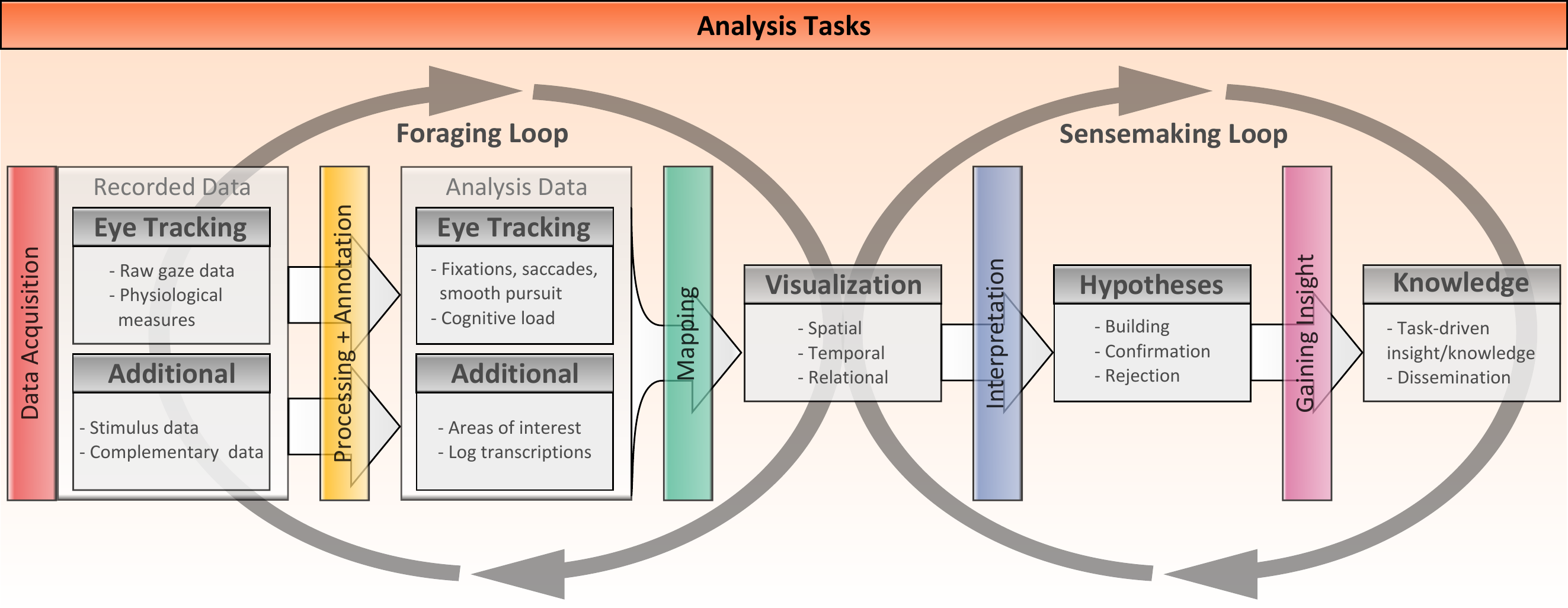}
    \caption{\label{fig:pipeline_eyetracking} 
    Schematic pipeline for the visual analysis of eye tracking data. All stages (data acquisition, processing, mapping, interpretation, gaining insight) are  influenced by the analysis task. 
    Figure reprinted by permission from Springer: book chapter by Kurzhals et al.~\cite{Kurzhals:2016:TBV} \copyright{} 2017.}
\end{figure}

Now, let us focus on eye tracking in empirical visualization research. Extending the fundamental visualization pipeline~\cite{Chi:2000:TVT,Haber:1990:VIC}, the process of acquisition and visual analysis of eye tracking data can be described by the pipeline of Figure~\ref{fig:pipeline_eyetracking}, as defined by Kurzhals et al.~\cite{Kurzhals:2016:TBV}. The study data consists of gaze information and---potentially---further complementary data. These are processed and annotated before the mapping to the visualization is computed. The overarching process consists of two interlinked loops: 
a foraging loop to investigate and explore the study observables,  and a sensemaking loop for the interpretation of the data~\cite{Pirolli:2005:SPV}. This interpretation may lead to confirming, rejecting, or building new hypotheses.

Figure~\ref{fig:pipeline_eyetracking} shows that data-rich information from eye tracking leads to a quite complex data analysis problem. General, rather high-level analysis tasks include compare, relate, and detect \cite{Kurzhals:2016:TBV}. There are a number of specific questions such as: on which parameters or data are these tasks performed (independent or dependent variables), do we want to define derived variables from raw data (other types of independent or dependent variables), which visualization techniques support these tasks and data types, what are the eventual research questions that should be answered by the analysis?

There is a comprehensive overview of visualization techniques for eye tracking data \cite{Blascheck:14:SAV,Blascheck:2017:VET}, along with a taxonomy that incorporates types of data, stimuli, and visualization techniques. Alternatively, Andrienko et al.~\cite{Andrienko:2012:VAM} provide a critical assessment and review of geo-inspired visual analytics techniques from the perspective of eye tracking analysis. These overview and review papers are a good starting point for choosing appropriate visualization techniques, depending on the visual analysis problem; see center part of Figure~\ref{fig:pipeline_eyetracking}. 

\begin{figure}[t]
    \centering
        \includegraphics[width=0.95\textwidth]{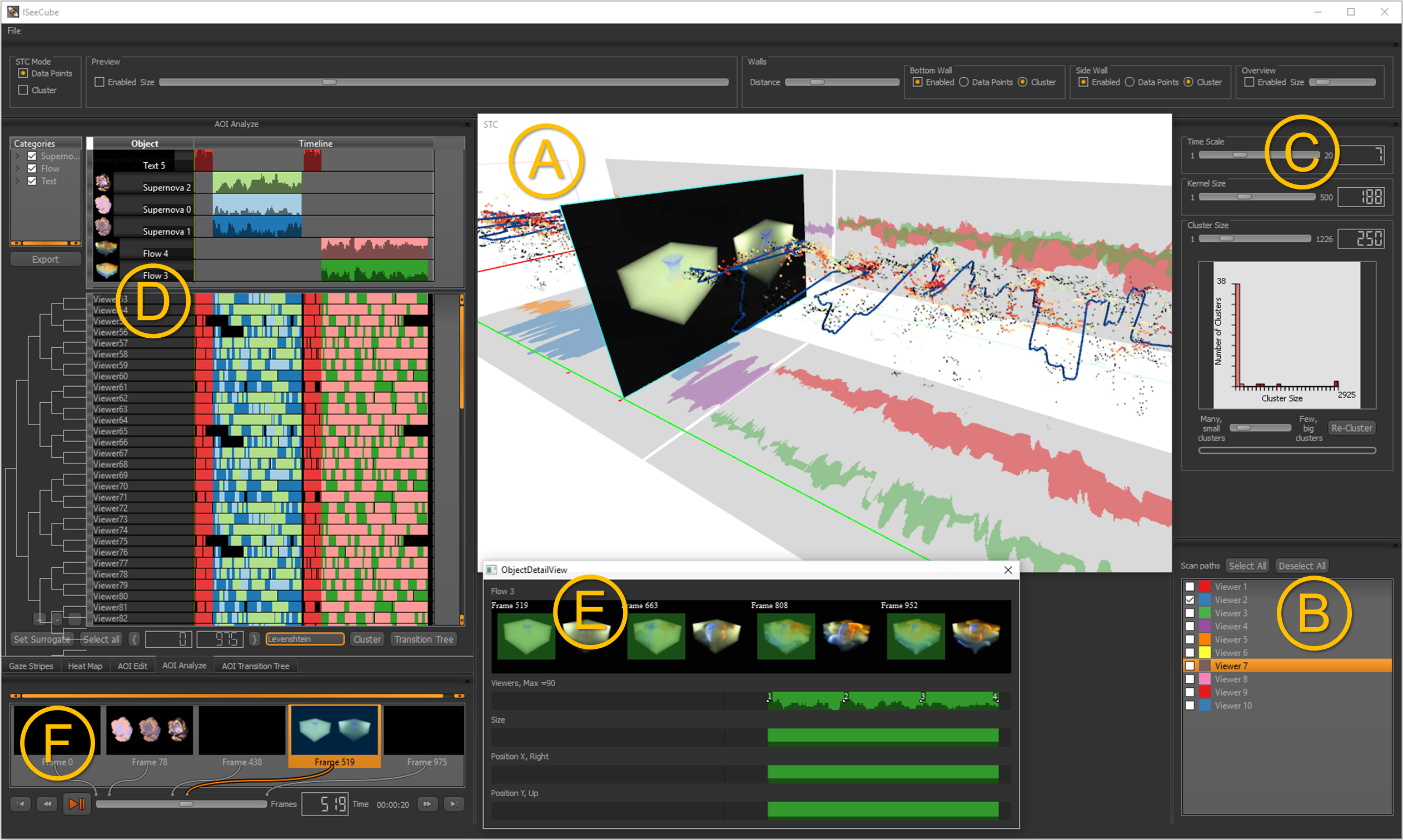}
    \caption{\label{fig:iseecube} Screenshot of the {\em ISeeCube} system~\cite{Kurzhals:2014:ISC}, which combines visual spatiotemporal gaze analysis with AOI-oriented analysis. The spatiotemporal analysis is based on a space-time cube visualization (A) that includes selected scanpaths (B) and the results of clustering controlled by user-specified parameters (C).  The AOI-oriented analysis is supported by hierarchical clustering and scarfplots of AOI sequences (D) and a detailed view of a selected AOI (F). The timeline of the video stimulus allows for temporal navigation (F).    
    The screenshot was taken when using {\em ISeeCube}~\cite{Kurzhals:2014:ISC} implemented by Kurzhals. Image \copyright{} 2019 Daniel Weiskopf}
\end{figure}

Overall, there has been quite some progress recently in novel and improved visualization techniques to support the evaluation of eye tracking studies. In particular, there are techniques that allow researchers to combine spatiotemporal gaze analysis~\cite{Kurzhals:2013:STV} with the integrated interpretation of scanpaths and areas of interest (AOIs)~\cite{Kurzhals:2014:ISC} (see Figure~\ref{fig:iseecube} for an example), visually compare scanpaths~\cite{Koch:2018:IBS}, examine large sets of gaze trajectories by bundling~\cite{Hurter:2013:BVD}, analyze time-dependent AOIs for long-timespan studies \cite{Muthumanickam:2019:ITV}, work with fixation metrics for the large-scale analysis of information visualizations \cite{Bylinskii:2015:EFM}, show gaze and stimulus simultaneously in a volume representation~\cite{Bruder:2019:STV}, or
relate gaze to data of interest in a visualization~\cite{Jianu:2018:DMT}.

There are many examples of the usefulness of such visual data analysis for eye tracking experiments. Typically, visual data analysis is a critical component in pilot studies that can then inform the design of the study process and statistical evaluation. I just want to briefly sketch a few typical examples of how visualization supported our own previous work on eye tracking evaluation of visualization techniques. One example is an eye tracking study that compares parallel coordinates and scatterplots \cite{Netzel:2017:CET}. Here, the visualization of scanpaths, attention, and AOIs for pilot studies helped us formulate hypotheses that eventually led to an advanced computational description of transitions between AOIs that could be used for statistical testing of complex reading behavior. Similarly, for an eye tracking study on transportation maps~\cite{Netzel:2017:UPR}, visualization allowed us to define a new numerical indicator for geodesic distance plots that served as a basis for statistical inference on reading behaviors. Finally, Burch et al.~\cite{Burch:2013:VTS} showcased many different types of visualization techniques and discussed how they could be used to identify qualitative findings in eye tracking data from a study on tree visualization techniques~\cite{Burch:2011:ETO}: visualization allowed us to identify reading strategies, reasons for the bad performance of radial tree layouts, and spatiotemporal characteristics of the eye tracking information. 

Despite the advances in visual analysis and the above success stories, Kurzhals et al.~\cite{Kurzhals:2016:ETE} pointed out a number of open issues related to evaluating visualization and visual analytics with eye tracking: we are still missing sufficient methods for scanpath comparison, fusion of different data sources (e.g., gaze with interaction logs or EEG), and practical tools and working analysis systems. Furthermore, Kurzhals et al.\ see the need of linking to cognitive models and translational evaluation of human cognition, which asks for building an interdisciplinary community that combines expertise in computer, cognitive, and social sciences. I think that these issues still remain as challenges today. In particular, the combination of data from different sources is a key aspect that needs to be addressed further. There is a need to reach out beyond eye tracking alone and include various other types of data that we can access during studies. 

Another challenge is scalability, especially if we want to address long-timespan studies and/or studies with large numbers of participants, leading to a big data visual analytics problem for eye tracking \cite{Blascheck:15:CPB}. This problem will also arise when visualization is evaluated with pervasive eye tracking \cite{Chuang:2019:UGS}, unconstrained mobile eye tracking, or in-the-wild research, typically with mobile eye tracking glasses. The analysis becomes challenging here because each study participant will see individual stimuli, which makes it hard to register or align gaze data between participants and relate them to the semantics of objects from the stimuli. In fact, the data analysis has to include much analysis for time-varying image data acquired by the world camera of the eye tracking glasses. There are some first attempts in this direction \cite{Kurzhals:2017:VAM} that combine computer-based image analysis with visual interaction, but we are still far from a simple, reliable, and time-efficient analysis process.


Up to now, the discussion has focused on eye tracking as an element of methods for quantitative research. However, for a more comprehensive evaluation approach, qualitative methods should also be considered---typically leading to a combination in the form of mixed methods \cite{Johnson:2007:TDM}. I see an integration of data-rich research methods (often the quantitative ones, especially when based on physiological sensors like eye tracking) with data-poor research methods (often the qualitative ones) as another area where visualization can play an important role.  An example of this research direction is the triangulation of different approaches (here, gaze combined with think-aloud protocol analysis and interaction logs) by Blascheck et al.~\cite{Blascheck2016:VA2,Blascheck:2016:TUB}. Taking this approach further, visual analysis and coding of participants' behavior and actions are possible~\cite{Blascheck:2016:VAC}, integrating data-rich gaze information in the form of word-sized graphics~\cite{Beck:2015:EWS} with other sources of information from experiments.

\section{Generalized Problem Characterization}
\label{sec:generalized}

The above discussion was centered around the specific example of eye tracking studies and the evaluation of the results of such studies. Many of the basic challenges already occur in this context of eye tracking and carry over to other types of studies. This section extends the discussion to a generalized view on visual data analysis for empirical visualization research.

\subsection{Data and Visualization Types}

The choice of visualization technique largely depends on the type of data that needs to be analyzed. In general, observational data will be large, complex, time-dependent, heterogeneous, and unstructured, coming from different types of sensors or information sources. However, in general, we can assume that observational data can be assigned some time stamp, i.e., data even from different sources can be eventually registered along the timeline (even though it might be difficult technically). In other words, the underlying data model is that of a time-dependent data set with different types of time-varying data attributes. 

The actual data attributes can be of largely varying type, and they may not be sampled at the same timepoints or same frequency. Some might not even be sampled at points in time, but spread across the timeline or even be associated with the full trial (i.e., the full timeline). There is a large set of potential variables that could be acquired as raw data during the experiments-
Typical types of time-series data consist of multidimensional data, i.e., multiple real-valued fields, or multiple categorical (nominal) data attributes (e.g., categories of events from user logs). Other types of much larger data sources include videos (images) and audio that may, for example, be recorded for protocol analysis or mobile eye tracking. Data may also include information about technical or algorithmic measures of performance \cite{Bruder:2019:ERP,Larsen:2016:PMI,Rizzi:2014:PMV}.
For any kind of such data, we may also obtain measures of reliability or uncertainty, which is relevant for many types of sensor data. 

The characterization of data does not stop at the stage of the original or raw data. In fact, many examples of visual analysis work on derived data that might be more informative than raw data. For the example of eye tracking in Figure~\ref{fig:pipeline_eyetracking}, the `analysis data' is typically derived data. Preferably, the derived data is fully automatically computed from the original sources, but there might be cases where user intervention might be required, for example, for the visual-interactive annotation of data. 

The choice of visualization technique(s) depends on the type of data to be analyzed. A general strategy is to use multiple coordinated views to support several data attributes~\cite{Roberts:2007:CMV}. More integrated visual representations may lead to better results but typically require a specific visual design. To address the complexity of the data analysis problem and facilitate scalability to large data, interactive visualization is routinely combined with automatic data analysis---such as statistical methods, unsupervised, or supervised learning---in a visual analytics setup. Finally, the choice of visualization may also depend on the independent variables, for example, whether we have to analyze data for individual participants or groups of participates, or whether we need comparative visualization to show differences with respect to independent variables.

\subsection{Analysis and Dissemination Goals}

Of course, the choice of visualization technique also depends on the goals of the analysis. Typical analysis tasks include outlier detection, summarization, aggregation, or grouping. A related perspective on data analysis goals for knowledge discovery in databases (KDD) is provided by Fayyad et al.~\cite{Fayyad:1996:KDD}.  Where possible, automatic data analysis or statistical techniques are employed to support the task, but as discussed above, the typical approach will follow the combination with interactive visualization. In particular, the visual analysis should also include the original input data or stimuli. The analysis of qualitative aspects of studies is especially challenging~\cite{Corbin:2015:BQR}; a general approach is based on coding such qualitative study data \cite{Saldana:2015:CMQ}. 

A fundamental issue of any visual data analysis is the question of reliability: interactive data exploration might lead to different findings, depending on the interaction steps taken by the analyst. This issue is present for the analysis of study data as well; after all, we want reliable and robust results from studies. Therefore, interactive visualization is typically accompanied by statistical analysis to obtain more controlled answers, yet based on hypotheses informed by visualization.
The sensemaking loop of Figure~\ref{fig:pipeline_eyetracking} indicates hypothesis building and testing for the example of eye tracking experiments; however, the general structure of the sensemaking loop extends to any kind of experimental evaluation and could include statistical testing.


Another issue is related to properly planning the setup of the studies. Their quality critically depends on an appropriate choice of stimuli or other input shown to the participants. Therefore, the generation of input data is of high relevance to support informative results of studies or facilitate benchmarking. A promising approach employs generative data models to do so \cite{Schulz:2016:GDM}. 

Finally, the goal of visualization does not stop at data analysis. In fact, visualization is equally relevant for disseminating results of studies after interpretation and insight generation in the sensemaking loop of Figure~\ref{fig:pipeline_eyetracking}. Therefore, visualization approaches for dissemination \cite{Beck:2016:VAD} and storytelling \cite{Lee:2015:MTT,Ma:2012:SSU} are required.

\section{Future Research Perspectives and Call for Action}
\label{sec:action}

Based on the specific observations and experiences with eye-tracking-based empirical visualization research (Section~\ref{sec:eyetracking}) and the generalized problem characterization (Section~\ref{sec:generalized}), I have identified the following, quite subjective recommendations for future research directions and a call for action.

\leftbar
\mbox{}\\[1ex]
\bf \mbox{}~~~Let us be our own domain experts: visualization for visualization ({\em Vis4Vis})!\\[0ex]
\mbox{}
\endleftbar

\noindent
I argue that we should prominently position visualization research as an application domain for visualization. So far, the call for papers and keywords in the paper submission systems of the main conferences of the visualization community (IEEE VIS, EuroVis, PacificVis) specifically ask for application or design study papers, but they do not explicitly consider visualization research---even in cases where they list many other research areas. Furthermore, the call for papers and submission keywords typically contain empirical research, especially user studies, but they focus on actual studies and not on methods that support the evaluation of studies. The series of BELIV Workshops is a good example of a venue that specifically asks for the development of research methods and, thus, implicitly supports the topic of {\em Vis4Vis}. Similarly, the series of Workshops on Eye Tracking and Visualization (ETVIS)\footnote{ETVIS: Workshop on Eye Tracking and Visualization. \url{https://www.etvis.org}} \cite{Burch:2016:ETV} facilitates such research, yet restricted to eye tracking. 

To advance our field, a more prominent integration of {\em Vis4Vis} in the main conferences would be helpful. 
%
Being our own domain experts has several benefits. First, we have an intrinsic and tight link to assessing whether our visual data analysis methods work well or how they need to be improved, leading to short development cycles; therefore, we can expect a fast development of useful visualization techniques that may even carry over to applications beyond those for empirical visualization research. Second, we will benefit from improved ways of evaluating our empirical studies, leading to a better understanding of visualization. Finally, since other disciplines such as HCI are facing similar evaluation challenges, there is a potential impact of improved data analysis for empirical research outside the visualization community.

\leftbar
\mbox{}\\[1ex]
\bf \mbox{}~~~Data-driven research for the next generation of empirical studies \\
\mbox{}~~~in visualization!\\[0ex]
\mbox{}
\endleftbar

\noindent
I am convinced that the integration of as-much-as-possible data acquired during studies is a viable way to conduct advanced empirical visualization studies that may support in-the-wild experiments, unconstrained settings, and individual participants and group work alike. Therefore, in the sense of {\em Vis4Vis}, we are facing the challenge of data fusion and combined visual analysis of massive, often messy sensor and other study data. This, in particular, may include various kinds of physiological sensor, image/video, and audio data. However, with the recent progress in machine learning, especially deep neural networks, there is a great potential that we will be able to work with data-rich experiments, with a strong emphasis on data-driven research. In fact, the combination of machine learning with visual analytics is a most promising approach to address these hard analysis problems, for example, in combination with video visual analytics \cite{Hoeferlin:2015:SVV}. In this context, it will be critical to keep the original data as long as possible in the analysis pipeline in order to be able to obtain reliable results. Furthermore, it is equally important to obtain reliable and controlled results for data analysis by complementing visual analysis with rigorous statistical testing.

\leftbar
\mbox{}\\[1ex]
\bf \mbox{}~~~New ways of reporting, privacy preservation, and open science!\\[0ex]
\mbox{}
\endleftbar

\noindent
With extended or new approaches to visual data analysis, we are also facing the issue of how we can report findings from empirical research. One part of this issue is the concise presentation of results, for example, in a research article. Here, traditional styles of reporting by using established statistical descriptions no longer work, but it is not yet clear how the wide variety of more complex analysis results could be summarized in a brief, yet comprehensible and replicable way. Here, visualization can play an important role in the sense of using it for storytelling of the scientific data, but respective methods are yet to be developed.

Another part of this issue is related to how we should communicate the massive data potentially acquired during studies. The straightforward approach is to provide the complete set of research data along with the publication, for example, in repositories that guarantee reliable and long-term access of open research data. However, raw data alone is not useful, and even if meta information is provided, it might still be hard to fully replicate previous studies if they come with complex data. Therefore, it might become relevant to even provide visual analysis tools and descriptions thereof along with the research data. Alternatively, our community could establish a set of tools on which the reproducibility of studies could rely, adopting similar ideas from eye tracking research~\cite{Munz:2019:VMV}. The issues of both storytelling and open science are connected to the development of visual data analysis methods in the sense of {\em Vis4Vis}. 

Furthermore, with open empirical data, we have to carefully consider issues related to privacy of participants and research ethics. With data-rich empirical data combined from different types of sensors, we might acquire enough information that could lead to a breach of anonymity if the data is published in original raw format, i.e., there is an intrinsic conflict between open science and privacy preservation. However, visualization has the potential to help here if it is extended toward novel privacy-preserving visualizations integrated into the research process. The outcome could be privacy-preserving, modified versions of the original data that could still be shared as open research data---with sufficient details to support reproducibility of the relevant research results.

\leftbar
\mbox{}\\[1ex]
\bf \mbox{}~~~Best practices for the next generation of evaluation methods!\\[0ex]
\mbox{}
\endleftbar

\noindent
The three areas of recommendations and future research directions mentioned above will have to be complemented by the adoption of the visualization techniques in the processes and reporting of empirical visualization research. To this end, I see an ongoing process of identifying best practices for novel evaluation approaches and establishing new standards of empirical research. 

\section{Conclusion}

It is obvious that visualization for visualization ({\em Vis4Vis}) is not the only answer to the challenges that we are facing in improving our set of methods for empirical visualization research. 
However, I am convinced that there is room for more advanced visualization methods for data analysis and reporting to be used in the context of studies within the visualization community, eventually improving our approach to empirical research.

\section*{Acknowledgments}

Funded by the Deutsche Forschungsgemeinschaft (DFG, German Research Foundation) -- Project-ID 251654672 -- TRR 161 (Project B01 and Task Force TF-B). I thank the participants of the Dagstuhl Seminar 18041 (``Foundations of Data Visualization'') for fruitful discussions. Special thanks to Kuno Kurzhals for the many discussions on eye tracking and visualization. The screenshot in Figure~\ref{fig:iseecube} was taken from his {\em ISeeCube} implementation \cite{Kurzhals:2014:ISC}.

\bibliographystyle{spmpsci}
\bibliography{ms}

\begin{thebibliography}{10}
\providecommand{\url}[1]{{#1}}
\providecommand{\urlprefix}{URL }
\expandafter\ifx\csname urlstyle\endcsname\relax
  \providecommand{\doi}[1]{DOI~\discretionary{}{}{}#1}\else
  \providecommand{\doi}{DOI~\discretionary{}{}{}\begingroup
  \urlstyle{rm}\Url}\fi

\bibitem{Anderson:12:EVU}
Anderson, E.W.: Evaluating visualization using cognitive measures.
\newblock In: Proceedings of the Workshop on Beyond Time And Errors: Novel
  Evaluation Methods for Visualization (BELIV), pp. 1--4 (2012)

\bibitem{Anderson:2011:USV}
Anderson, E.W., Potter, K.C., Matzen, L.E., Shepherd, J.F., Preston, G.A.,
  Silva, C.T.: A user study of visualization effectiveness using {EEG} and
  cognitive load.
\newblock Computer Graphics Forum \textbf{30}(3), 791--800 (2011)

\bibitem{Andrienko:2012:VAM}
Andrienko, G.L., Andrienko, N.V., Burch, M., Weiskopf, D.: Visual analytics
  methodology for eye movement studies.
\newblock IEEE Transactions on Visualization and Computer Graphics
  \textbf{18}(12), 2889--2898 (2012)

\bibitem{Beck:2015:EWS}
Beck, F., Blascheck, T., Ertl, T., Weiskopf, D.: Exploring word-sized graphics
  for visualizing eye tracking data within transcribed experiment recordings.
\newblock In: M.~Burch, L.~Chuang, B.~Fisher, A.~Schmidt, D.~Weiskopf (eds.)
  Eye Tracking and Visualization: Foundations, Techniques, and Applications,
  pp. 113--128. Springer (2016)

\bibitem{Beck:2016:VAD}
Beck, F., Koch, S., Weiskopf, D.: Visual analysis and dissemination of
  scientific literature collections with {SurVis}.
\newblock IEEE Transactions on Visualization and Computer Graphics
  \textbf{22}(1), 180--189 (2016)

\bibitem{Blascheck:2016:VAC}
Blascheck, T., Beck, F., Baltes, S., Ertl, T., Weiskopf, D.: Visual analysis
  and coding of data-rich user behavior.
\newblock In: Proceedings of the {IEEE} Conference on Visual Analytics Science
  and Technology, pp. 141--150 (2016)

\bibitem{Blascheck:15:CPB}
Blascheck, T., Burch, M., Raschke, M., Weiskopf, D.: Challenges and
  perspectives in big eye-movement data visual analytics.
\newblock In: Proceedings of the IEEE International Symposium on Big Data
  Visual Analytics, pp. 1--8 (2015)

\bibitem{Blascheck:2014:TAE}
Blascheck, T., Ertl, T.: Towards analyzing eye tracking data for evaluating
  interactive visualization systems.
\newblock In: Proceedings of the Workshop on Beyond Time And Errors: Novel
  Evaluation Methods for Visualization (BELIV), pp. 70--77 (2014)

\bibitem{Blascheck:2016:TUB}
Blascheck, T., John, M., Koch, S., Bruder, L., Ertl, T.: Triangulating user
  behavior using eye movement, interaction, and think aloud data.
\newblock In: Proceedings of the ACM Symposium on Eye Tracking Research \&
  Applications, pp. 175--182 (2016)

\bibitem{Blascheck2016:VA2}
Blascheck, T., John, M., Kurzhals, K., Koch, S., Ertl, T.: {VA$^2$: A} visual
  analytics approach for evaluating visual analytics applications.
\newblock IEEE Transactions on Visualization and Computer Graphics \textbf{22},
  61--70 (2016)

\bibitem{Blascheck:14:SAV}
Blascheck, T., Kurzhals, K., Raschke, M., Burch, M., Weiskopf, D., Ertl, T.:
  State-of-the-art of visualization for eye tracking data.
\newblock In: EuroVis -- STARs, pp. 63--82 (2014)

\bibitem{Blascheck:2017:VET}
Blascheck, T., Kurzhals, K., Raschke, M., Burch, M., Weiskopf, D., Ertl, T.:
  Visualization of eye tracking data: {A} taxonomy and survey.
\newblock Computer Graphics Forum \textbf{36}(8), 260--284 (2017)

\bibitem{Bruder:2019:STV}
Bruder, V., Kurzhals, K., Frey, S., Weiskopf, D., Ertl, T.: Space-time volume
  visualization of gaze and stimulus.
\newblock In: Proceedings of the ACM Symposium on Eye Tracking Research {\&}
  Applications, pp. 12:1--12:9 (2019)

\bibitem{Bruder:2019:ERP}
Bruder, V., M{\"u}ller, C., Frey, S., Ertl, T.: On evaluating runtime
  performance of interactive visualizations.
\newblock IEEE Transactions on Visualization and Computer Graphics  (2019).
\newblock \doi{10.1109/TVCG.2019.2898435}

\bibitem{Bulling:2011:EMA}
Bulling, A., Ward, J.A., Gellersen, H., Troster, G.: Eye movement analysis for
  activity recognition using electrooculography.
\newblock {IEEE} Transactions on Pattern Analysis and Machine Intelligence
  \textbf{33}(4), 741--753 (2011)

\bibitem{Burch:2013:VTS}
Burch, M., Andrienko, G.L., Andrienko, N.V., H{\"{o}}ferlin, M., Raschke, M.,
  Weiskopf, D.: Visual task solution strategies in tree diagrams.
\newblock In: Proceedings of the {IEEE} Pacific Visualization Symposium, pp.
  169--176 (2013)

\bibitem{Burch:2016:ETV}
Burch, M., Chuang, L., Fisher, B., Schmidt, A., Weiskopf, D. (eds.): Eye
  Tracking and Visualization: Foundations, Techniques, and Applications.
\newblock Springer (2016)

\bibitem{Burch:2011:ETO}
Burch, M., Konevtsova, N., Heinrich, J., H{\"{o}}ferlin, M., Weiskopf, D.:
  Evaluation of traditional, orthogonal, and radial tree diagrams by an eye
  tracking study.
\newblock IEEE Transactions on Visualization and Computer Graphics
  \textbf{17}(12), 2440--2448 (2011)

\bibitem{Bylinskii:2015:EFM}
Bylinskii, Z., Borkin, M.A.: Eye fixation metrics for large scale analysis of
  information visualizations.
\newblock In: M.~Burch, L.~Chuang, B.~Fisher, A.~Schmidt, D.~Weiskopf (eds.)
  Eye Tracking and Visualization: Foundations, Techniques, and Applications,
  pp. 235--255. Springer (2016)

\bibitem{Carpendale:08:EIV}
Carpendale, S.: Evaluating information visualizations.
\newblock In: A.~Kerren, J.T. Stasko, J.D. Fekete, C.~North (eds.) Information
  Visualization: Human-Centered Issues and Perspectives, pp. 19--45. Springer
  (2008)

\bibitem{Chi:2000:TVT}
Chi, E.H.: A taxonomy of visualization techniques using the data state
  reference model.
\newblock In: Proceedings of the IEEE Symposium on Information Visualization,
  pp. 69--75 (2000)

\bibitem{Chuang:2019:UGS}
Chuang, L., Duchowski, A., Qvarfordt, P., Weiskopf, D.: Ubiquitous gaze sensing
  and interaction ({Dagstuhl Seminar} 18252).
\newblock Dagstuhl Reports \textbf{8}(6), 77--148 (2019)

\bibitem{Coltekin:2009:EEI}
{\c{C}}{\"o}ltekin, A., Heil, B., Garlandini, S., Fabrikant, S.I.: Evaluating
  the effectiveness of interactive map interface designs: a case study
  integrating usability metrics with eye-movement analysis.
\newblock Cartography and Geographic Information Science \textbf{36}(1), 5--17
  (2009)

\bibitem{Corbin:2015:BQR}
Corbin, J., Strauss, A.: Basics of Qualitative Research: Techniques and
  Procedures for Developing Grounded Theory, 4th edn.
\newblock SAGE Publications (2015)

\bibitem{Crabtree:2013:ISI}
Crabtree, A., Chamberlain, A., Grinter, R.E., Jones, M., Rodden, T., Rogers,
  Y.: Introduction to the special issue of `the turn to the wild'.
\newblock ACM Transactions on Computer-Human Interaction \textbf{20}(3),
  13:1--13:4 (2013)

\bibitem{Cui:2011:QCN}
Cui, X., Bray, S., Bryant, D.M., Glover, G.H., Reiss, A.L.: A quantitative
  comparison of {NIRS} and {fMRI} across multiple cognitive tasks.
\newblock Neuroimage \textbf{54}(4), 2808--2821 (2011)

\bibitem{Duchowski:2007:ETM}
Duchowski, A.: Eye Tracking Methodology: Theory and Practice, 2nd edn.
\newblock Springer (2007)

\bibitem{Ellis:06:EAU}
Ellis, G., Dix, A.J.: An explorative analysis of user evaluation studies in
  information visualisation.
\newblock In: Proceedings of the Workshop on Beyond Time And Errors: Novel
  Evaluation Methods for Visualization (BELIV), pp. 1--7 (2006)

\bibitem{Elmqvist:12:PVE}
Elmqvist, N., Yi, J.S.: Patterns for visualization evaluation.
\newblock In: Proceedings of the Workshop on Beyond Time And Errors: Novel
  Evaluation Methods for Visualization (BELIV), pp. 12:1--12:8 (2012)

\bibitem{Ericsson:1993:PAV}
Ericsson, K.A., Simon, H.A.: Protocol Analysis: Verbal Reports as Data, revised
  edn.
\newblock MIT Press (1993)

\bibitem{Fathi:2012:LRD}
Fathi, A., Li, Y., Rehg, J.M.: Learning to recognize daily actions using gaze.
\newblock In: Proceedings of the European Conference on Computer Vision, pp.
  314--327. Springer (2012)

\bibitem{Fayyad:1996:KDD}
Fayyad, U., Piatetsky-Shapiro, G., Smyth, P.: The {KDD} process for extracting
  useful knowledge from volumes of data.
\newblock Communications of the ACM \textbf{39}(11), 27--34 (1996)

\bibitem{Freitas:14:HIC}
Freitas, C.M.D.S., Pimenta, M.S., Scapin, D.L.: User-centered evaluation of
  information visualization techniques: Making the {HCI-InfoVis} connection
  explicit.
\newblock In: W.~Huang (ed.) Handbook of Human Centric Visualization, pp.
  315--336. Springer (2014)

\bibitem{Goldberg:10:CIG}
Goldberg, J.H., Helfman, J.I.: Comparing information graphics: a critical look
  at eye tracking.
\newblock In: Proceedings of the Workshop on Beyond Time And Errors: Novel
  Evaluation Methods for Visualization (BELIV), pp. 71--78 (2010)

\bibitem{Gurkok:2012:BCI}
G{\"u}rk{\"o}k, H., Nijholt, A.: Brain--computer interfaces for multimodal
  interaction: a survey and principles.
\newblock International Journal of Human-Computer Interaction \textbf{28}(5),
  292--307 (2012)

\bibitem{Haber:1990:VIC}
Haber, R.B., McNabb, D.A.: Visualization idioms: A conceptual model for
  visualization systems.
\newblock In: G.M. Nielson, B.D. Shriver, L.J. Rosenblum (eds.) Visualization
  in Scientific Computing, pp. 74--93. IEEE Computer Society Press (1990)

\bibitem{Hirshfield:2011:TYB}
Hirshfield, L.M., Gulotta, R., Hirshfield, S., Hincks, S., Russell, M., Ward,
  R., Williams, T., Jacob, R.: This is your brain on interfaces: enhancing
  usability testing with functional near-infrared spectroscopy.
\newblock In: Proceedings of the {SIGCHI} Conference on Human Factors in
  Computing Systems, pp. 373--382 (2011)

\bibitem{Hoeferlin:2015:SVV}
H{\"{o}}ferlin, B., H{\"{o}}ferlin, M., Heidemann, G., Weiskopf, D.: Scalable
  video visual analytics.
\newblock Information Visualization \textbf{14}(1), 10--26 (2015)

\bibitem{Holmqvist:2011:ETC}
Holmqvist, K., Nystr{\"o}m, M., Andersson, R., Dewhurst, R., Jarodzka, H.,
  Van~de Weijer, J.: Eye Tracking: A Comprehensive Guide to Methods and
  Measures.
\newblock Oxford University Press (2011)

\bibitem{Hurter:2013:BVD}
Hurter, C., Ersoy, O., Fabrikant, S., Klein, T., Telea, A.: Bundled
  visualization of dynamic graph and trail data.
\newblock IEEE Transactions on Visualization and Computer Graphics
  \textbf{20}(8), 1141--1157 (2013)

\bibitem{Hutchins:1995:CW}
Hutchins, E.: Cognition in the Wild.
\newblock MIT Press (1995)

\bibitem{Isenberg:2013:SRP}
Isenberg, T., Isenberg, P., Chen, J., Sedlmair, M., M{\"{o}}ller, T.: A
  systematic review on the practice of evaluating visualization.
\newblock IEEE Transactions on Visualization and Computer Graphics
  \textbf{19}(12), 2818--2827 (2013)

\bibitem{Jacob:2003:ETH}
Jacob, R.J., Karn, K.S.: Eye tracking in human-computer interaction and
  usability research: Ready to deliver the promises.
\newblock In: J.~Hy{\"o}n{\"a}, R.~Radach, H.~Deubel (eds.) The Mind's Eye:
  Cognitive and Applied Aspects of Eye Movement Research, pp. 573--605.
  Elsevier (2003)

\bibitem{Jianu:2018:DMT}
Jianu, R., Alam, S.S.: A data model and task space for data of interest {(DOI)}
  eye-tracking analyses.
\newblock IEEE Transactions on Visualization and Computer Graphics
  \textbf{24}(3), 1232--1245 (2018)

\bibitem{Johnson:2007:TDM}
Johnson, R.B., Onwuegbuzie, A.J., Turner, L.A.: Toward a definition of mixed
  methods research.
\newblock Journal of Mixed Methods Research \textbf{1}(2), 112--133 (2007)

\bibitem{Kim:2012:DET}
Kim, S.H., Dong, Z., Xian, H., Upatising, B., Yi, J.S.: Does an eye tracker
  tell the truth about visualizations?: Findings while investigating
  visualizations for decision making.
\newblock IEEE Transactions Visualization Computer Graphics \textbf{18}(12),
  2421--2430 (2012)

\bibitem{Koch:2018:IBS}
Koch, M., Kurzhals, K., Weiskopf, D.: Image-based scanpath comparison with
  slit-scan visualization.
\newblock In: Proceedings of the ACM Symposium on Eye Tracking Research \&
  Applications, pp. 55:1--55:5 (2018)

\bibitem{Kurzhals:2016:TBV}
Kurzhals, K., Burch, M., Blascheck, T., Andrienko, G., Andrienko, N., Weiskopf,
  D.: A task-based view on the visual analysis of eye-tracking data.
\newblock In: M.~Burch, L.~Chuang, B.~Fisher, A.~Schmidt, D.~Weiskopf (eds.)
  Eye Tracking and Visualization: Foundations, Techniques, and Applications,
  pp. 3--22. Springer (2016)

\bibitem{Kurzhals:2014:EVA}
Kurzhals, K., Fisher, B.D., Burch, M., Weiskopf, D.: Evaluating visual
  analytics with eye tracking.
\newblock In: Proceedings of the Workshop on Beyond Time And Errors: Novel
  Evaluation Methods for Visualization (BELIV), pp. 61--69 (2014)

\bibitem{Kurzhals:2016:ETE}
Kurzhals, K., Fisher, B.D., Burch, M., Weiskopf, D.: Eye tracking evaluation of
  visual analytics.
\newblock Information Visualization \textbf{15}(4), 340--358 (2016)

\bibitem{Kurzhals:2014:ISC}
Kurzhals, K., Heimerl, F., Weiskopf, D.: {ISeeCube}: Visual analysis of gaze
  data for video.
\newblock In: Proceedings of the ACM Symposium on Eye Tracking Research \&
  Applications, pp. 43--50 (2014)

\bibitem{Kurzhals:2017:VAM}
Kurzhals, K., Hlawatsch, M., Seeger, C., Weiskopf, D.: Visual analytics for
  mobile eye tracking.
\newblock {IEEE} Transactions on Visualization and Computer Graphics
  \textbf{23}(1), 301--310 (2017)

\bibitem{Kurzhals:2013:STV}
Kurzhals, K., Weiskopf, D.: Space-time visual analytics of eye-tracking data
  for dynamic stimuli.
\newblock IEEE Transactions on Visualization and Computer Graphics
  \textbf{19}(12), 2129--2138 (2013)

\bibitem{Lam:2012:ESI}
Lam, H., Bertini, E., Isenberg, P., Plaisant, C., Carpendale, S.: Empirical
  studies in information visualization: Seven scenarios.
\newblock IEEE Transactions on Visualization and Computer Graphics
  \textbf{18}(9), 1520--1536 (2012)

\bibitem{Lam:08:IUQ}
Lam, H., Munzner, T.: Increasing the utility of quantitative empirical studies
  for meta-analysis.
\newblock In: Proceedings of the Workshop on Beyond Time And Errors: Novel
  Evaluation Methods for Visualization (BELIV) (2008).
\newblock Article No. 2

\bibitem{Larsen:2016:PMI}
Larsen, M., Harrison, C., Kress, J., Pugmire, D., Meredith, J.S., Childs, H.:
  Performance modeling of in situ rendering.
\newblock In: Proceedings of the International Conference for High Performance
  Computing, Networking, Storage and Analysis, pp. 276--287 (2016)

\bibitem{Lave:1988:CP}
Lave, J.: Cognition in Practice.
\newblock Cambridge University Press (1988)

\bibitem{Lee:2015:MTT}
Lee, B., Henry~Riche, N., Isenberg, P., Carpendale, S.: More than telling a
  story: Transforming data into visually shared stories.
\newblock IEEE Computer Graphics and Applications \textbf{35}(5), 84--90 (2015)

\bibitem{Ma:2012:SSU}
Ma, K., Liao, I., Frazier, J., Hauser, H., Kostis, H.: Scientific storytelling
  using visualization.
\newblock IEEE Computer Graphics and Applications \textbf{32}(1), 12--19 (2012)

\bibitem{Marriott:2018:IA}
Marriott, K., Schreiber, F., Dwyer, T., Klein, K., Riche, N.H., Itoh, T.,
  Stuerzlinger, W., Thomas, B.H. (eds.): Immersive Analytics.
\newblock Springer (2018)

\bibitem{Munz:2019:VMV}
Munz, T., Chuang, L., Pannasch, S., Weiskopf, D.: {VisME:} visual microsaccades
  explorer.
\newblock Journal of Eye Movement Research \textbf{12}(6) (2019).
\newblock \doi{10.16910/jemr.12.6.5}

\bibitem{Muthumanickam:2019:ITV}
Muthumanickam, P.K., Vrotsou, K., Nordman, A., Johansson, J., Cooper, M.D.:
  Identification of temporally varying areas of interest in long-duration
  eye-tracking data sets.
\newblock IEEE Transactions on Visualization and Computer Graphics
  \textbf{25}(1), 87--97 (2019)

\bibitem{Netzel:2017:UPR}
Netzel, R., Ohlhausen, B., Kurzhals, K., Woods, R., Burch, M., Weiskopf, D.:
  User performance and reading strategies for metro maps: An eye tracking
  study.
\newblock Spatial Cognition {\&} Computation \textbf{17}(1-2), 39--64 (2017)

\bibitem{Netzel:2017:CET}
Netzel, R., Vuong, J., Engelke, U., O'Donoghue, S.I., Weiskopf, D., Heinrich,
  J.: Comparative eye-tracking evaluation of scatterplots and parallel
  coordinates.
\newblock Visual Informatics \textbf{1}(2), 118--131 (2017)

\bibitem{North:2006:TMV}
North, C.: Toward measuring visualization insight.
\newblock IEEE Computer Graphics and Applications \textbf{26}(3), 6--9 (2006)

\bibitem{Peck:2010:YBY}
Peck, E.M., Solovey, E.T., Chauncey, K., Sassaroli, A., Fantini, S., Jacob,
  R.J.K., Girouard, A., Hirshfield, L.M.: Your brain, your computer, and you.
\newblock Computer \textbf{43}(12), 86--89 (2010)

\bibitem{Pirolli:2005:SPV}
Pirolli, P., Card, S.: The sensemaking process and leverage points for analyst
  technology as identified through cognitive task analysis.
\newblock In: Proceedings of the International Conference on Intelligence
  Analysis, vol.~5, pp. 2--4 (2005)

\bibitem{Plaisant:2004:CIV}
Plaisant, C.: The challenge of information visualization evaluation.
\newblock In: Proceedings of the Working Conference on Advanced Visual
  Interfaces, pp. 109--116 (2004)

\bibitem{Prendinger:2005:UHP}
Prendinger, H., Mori, J., Ishizuka, M.: Using human physiology to evaluate
  subtle expressivity of a virtual quizmaster in a mathematical game.
\newblock International Journal of Human-Computer Studies \textbf{62}(2),
  231--245 (2005)

\bibitem{Rizzi:2014:PMV}
Rizzi, S., Hereld, M., Insley, J., Papka, M.E., Uram, T., Vishwanath, V.:
  Performance modeling of vl3 volume rendering on {GPU}-based clusters.
\newblock In: Proceedings of the Eurographics Symposium on Parallel Graphics
  and Visualization, pp. 65--72 (2014)

\bibitem{Roberts:2007:CMV}
Roberts, J.C.: State of the art: Coordinated multiple views in exploratory
  visualization.
\newblock In: Proceedings of the International Conference on Coordinated and
  Multiple Views in Exploratory Visualization, pp. 61--71 (2007)

\bibitem{Saldana:2015:CMQ}
Saldana, J.: The Coding Manual for Qualitative Researchers, 3rd edn.
\newblock SAGE Publications (2015)

\bibitem{Schulz:2016:GDM}
Schulz, C., Nocaj, A., El{-}Assady, M., Frey, S., Hlawatsch, M., Hund, M.,
  Karch, G.K., Netzel, R., Sch{\"{a}}tzle, C., Butt, M., Keim, D.A., Ertl, T.,
  Brandes, U., Weiskopf, D.: Generative data models for validation and
  evaluation of visualization techniques.
\newblock In: Proceedings of the Workshop on Beyond Time And Errors: Novel
  Evaluation Methods for Visualization (BELIV), pp. 112--124 (2016)

\bibitem{Shao:2017:VEL}
Shao, L., Silva, N., Eggeling, E., Schreck, T.: Visual exploration of large
  scatter plot matrices by pattern recommendation based on eye tracking.
\newblock In: Proceedings of the ACM Workshop on Exploratory Search and
  Interactive Data Analytics, pp. 9--16 (2017)

\bibitem{Silva:2019:ETS}
Silva, N., Blascheck, T., Jianu, R., Rodrigues, N., Weiskopf, D., Raubal, M.,
  Schreck, T.: Eye tracking support for visual analytics systems: foundations,
  current applications, and research challenges.
\newblock In: Proceedings of the ACM Symposium on Eye Tracking Research {\&}
  Applications, pp. 11:1--11:10 (2019)

\bibitem{Silva:2018:LEG}
Silva, N., Schreck, T., Veas, E., Sabol, V., Eggeling, E., Fellner, D.W.:
  Leveraging eye-gaze and time-series features to predict user interests and
  build a recommendation model for visual analysis.
\newblock In: Proceedings of the ACM Symposium on Eye Tracking Research \&
  Applications (2018).
\newblock Article No. 13

\bibitem{Steichen:2013:UAI}
Steichen, B., Carenini, G., Conati, C.: User-adaptive information
  visualization: using eye gaze data to infer visualization tasks and user
  cognitive abilities.
\newblock In: Proceedings of the {ACM} International Conference on Intelligent
  User Interfaces, pp. 317--328 (2013)

\bibitem{Strait:2014:RNB}
Strait, M., Canning, C., Scheutz, M.: Reliability of {NIRS-based BCIs}: A
  placebo-controlled replication and reanalysis of {Brainput}.
\newblock In: CHI '14 Extended Abstracts on Human Factors in Computing Systems,
  pp. 619--630 (2014)

\bibitem{Suchman:1987:PSA}
Suchman, L.A.: Plans and Situated Actions: The Problem of Human-Machine
  Communication.
\newblock Cambridge University Press (1987)

\bibitem{Tory:14:USV}
Tory, M.: User studies in visualization: A reflection on methods.
\newblock In: W.~Huang (ed.) Handbook of Human Centric Visualization, pp.
  411--426. Springer (2014)

\bibitem{Vuillemot:2016:LIV}
Vuillemot, R., Boy, J., Tabard, A., Perin, C., Fekete, J.D. (eds.): Proceedings
  of {LIVVIL:} Logging Interactive Visualizations and Visualizing Interaction
  Logs (2016).
\newblock Workshop at IEEE VIS 2016,
  \url{https://hal.inria.fr/hal-01535913/file/proceedings.pdf}

\bibitem{Wagner:2005:PSE}
Wagner, J., Kim, J., Andr{\'e}, E.: From physiological signals to emotions:
  Implementing and comparing selected methods for feature extraction and
  classification.
\newblock In: Proceedings of the {IEEE} International Conference on Multimedia
  and Expo, pp. 940--943 (2005)

\bibitem{Wijk:2013:ECV}
van Wijk, J.J.: Evaluation: A challenge for visual analytics.
\newblock IEEE Computer \textbf{46}(7), 56--60 (2013)

\end{thebibliography}

\end{document}